\documentstyle[aps,pre,epsf]{revtex} 

\newcommand{\nc}{\newcommand}
\nc{\beq}{\begin{equation}}
\nc{\eeq}{\end{equation}}
\nc{\bega}{\begin{eqnarray}}
\nc{\ega}{\end{eqnarray}}
\nc{\non}{\nonumber}
\nc{\vrho}{\varrho}
\nc{\ptx}{\partial_x}
\nc{\dt}{\partial_t} 
\nc{\rmd}{\rm d} 
\nc{\6}{\partial} 
\nc{\eq}[1]{(\ref{eq:#1})} 

\begin{document}
\date{\today}
\title{Modelling thermostatting, entropy currents and 
cross effects by dynamical systems}
\author{J\"urgen Vollmer$^{(1,2)}$, 
Tam\'as T\'el,$^{(3)}$ and
L\'aszl\'o M\'aty\'as$^{(3)}$}
\address{
(1) Fachbereich Physik, Univ.-GH Essen, 45117 Essen, Germany.
\\
(2) Max-Planck-Institut for Polymer Research,
   Ackermannweg 10, 
   55128 Mainz, Germany.
\\ 
(3) Institute for Theoretical Physics,
E\"otv\"os University,
P. O. Box 32,
H-1518 Budapest,
Hungary.
}
\maketitle
\begin{abstract}
A generalized multibaker map 
with periodic boundary conditions is shown to model boundary-driven 
transport, when the driving is applied by a ``perturbation'' of the dynamics
localised in a macroscopically small region. 
In this case there are sustained density gradients in the steady state. 
A non-uniform stationary temperature profile can be maintained 
by incorporating a heat source into the dynamics, which deviates from the 
one of a bulk system only in a (macroscopically small) localized
region such that a heat (or entropy) flux can enter an attached
thermostat only in that region. 
For these settings the relation between the average phase-space
contraction, the entropy flux to the thermostat and irreversible
entropy production is clarified for stationary and non-stationary
states. 
In addition, thermoelectric cross-effects are described by
a multibaker chain consisting of two parts with different
transport properties, modelling a junction between two metals. 
\end{abstract}

\section{Introduction}

There is a recent interest in modelling transport processes by 
simple dynamical systems with 
chaotic dynamics. 
One class of models, actually inspired by  
{\underline n}on-{\underline e}quilibrium 
{\underline m}olecular {\underline d}ynamics (NEMD) simulations, 
describes systems driven by external fields with a spatially
uniform dynamics subjected to periodic boundary conditions 
\cite{books,Vance,CELS,ECM,GC,KKN,Dorf}. 
Another approach concentrates on systems driven from the boundaries, 
which lead to steady states 
with sustained gradients of the thermodynamic fields 
\cite{CL,WKN}. 
For a comparatively simple, 
but as far as their transport properties are concerned, generic class 
of dynamical systems, the {\em multibakers,\/} 
\cite{G,TG,VTB97,Gent,VTB98,BTV98,GD,MTV99,MTV00,TG98,Gasp}  
we show that both mechanisms of driving can 
simultaneously be worked out. 
This leads to an improved understanding of the relation between the approaches. 
In the former approach transport is driven by a field 
acting uniformly in the full system, while in the latter case 
the driving is concentrated to a microscopic region in space. 
{F}rom this point of view, boundary-driven transport is closely analogous to 
transport in a dynamical system with periodic boundary conditions, 
which is driven out of equilibrium by a ``perturbation'' 
of the dynamics, localised in a macroscopically small region. 

In all models for transport, as emphasized by Nicolis and coworkers 
\cite{ND,DN}, 
a quantity of central interest is the heat flux, or equivalently the 
entropy flux, from the system into its environment. 
A central aim of modelling transport by dynamical systems 
is to identify settings, which 
are consistent with the thermodynamic entropy balance 
\begin{equation}
   \frac{{\rm d} S}{{\rm d} t}
 =
   \frac{{\rm d}_e S}{{\rm d} t} +
   \frac{{\rm d}_i S}{{\rm d} t} , 
\label{eq:entrbg}
\end{equation}
i.e., with the statement that the temporal change of the 
thermodynamic entropy $S$ can be decomposed 
into two contributions, called  
the external and internal change of the entropy, respectively.
This integral form can be rewritten into a local balance equation when 
the two terms on the right hand side correspond to integrals of local
densities. 
In that case, the time derivative of the entropy density $s$
appears as
\begin{equation}
   \dt s = \Phi + \sigma^{\rm (irr)} 
\label{eq:cs}
\end{equation}
where $\Phi$ and $\sigma^{(irr)}$ represent the densities
of the entropy flux and the rate of irreversible entropy
production, respectively.
In the bulk of typical macroscopic systems the entropy flux can 
be written as the divergence of the  entropy current $j^{(s)}$, 
\begin{equation}
   \Phi= - \nabla j^{(s)} , 
\label{Phith}
\end{equation}
reflecting the fact that no heat can be taken out from the 
system locally \cite{GM}. 
On the other hand, this form has to be generalized at positions where there 
is a heat current flowing into an attached thermostat, and in cases where 
the entropy current is not differentiable, 
like for istance across interfaces between different materials. 
In those cases the entropy flux is not a full divergence, and it need not even 
be defined as a density. 
Rather the flux should then be written as 
\begin{equation} \label{eq:balance0}
   \Phi= - \nabla j^{(s)}+ \Phi^{(th)} , 
\end{equation}
where $j^{(s)}$ is still the entropy current flowing in the system, 
but $\Phi^{(th)}$ accounts for 
the heat taken out in the form of a direct flow into the surroundings, which 
acts then as a thermostat.

In the present paper, we shall put special emphasis on the role of the entropy flux 
$\Phi^{(th)}$, and on exploring under which conditions it can vanish. 
We find conditions on how to model the entropy balance for thermodynamic 
bulk systems, 
and for macroscopic systems, which are subjected to thermostatting by either 
a localized sink for the entropy or a spatially uniform coupling 
to a thermostat.

The role of the thermodynamic entropy of dynamical systems 
is played by the coarse-grained 
Gibbs entropy, whose usefulness in understanding irreversibility 
from the point of view of 
dynamical systems is by now thoroughly explained in the literature 
\cite{Ru,BTV96,ND,DN,RM,Gal,VTB97,Gent,VTB98,BTV98,GD,MTV99,MTV00,TG98,Gasp}. 
({ For stochastically perturbed dynamical systems where noise generates 
a kind of coarse graining, see} \cite{ND,DN}). 
The bulk dynamics is represented by a multibaker model 
driving two fields, the density 
$\varrho$ and the the kinetic energy per particle $T$, 
with a local source density 
$q$ for the latter \cite{MTV99,MTV00}. 
A connection with macroscopic transport equations is aimed at in a suitable defined 
continuum limit (the {\em macroscopic limit\/}), where the field $T$ will be interpreted
as a temperature, based on the experience that this quantity is closely related to the 
average kinetic energy per particle.
 
We shall consider a sequence of periodic models of increasing complexity. 
Model I corresponds to a homogeneous isothermal system described by a
thermostatting algorithm. 
In this model no entropy current is defined --- its entire entropy flux stems from a $\Phi^{(th)}$. 
By allowing  a spatial resolution of the isothermal system (Model II),
a non-vanishing $- \nabla j^{(s)}$ term appears in the transient behaviour, but  $\Phi^{(th)}$ remains unchanged. 
It is the only contribution to the flux in a steady state. 
Model III is still isothermal but with a locally deviating dynamics in one of the multibaker cells representing a boundary. 
The bulk dynamics can then be chosen so that (\ref{Phith}) holds in the bulk, and all the heat taken out is concentrated in the boundary with a $\Phi^{(th)} \not= 0$ there.
In model IV we allow for temperature changes and local heat sources.
By taking $q$ locally deviating from that of the bulk in one cell,
we find a steady temperature profile with a break at the boundary. 
The $q$ distribution can then be chosen such that again (\ref{Phith}) 
holds in the bulk. 
The heat source in the boundary is however singular. 
It corresponds exactly to the one which follows from thermodynamics. 
Finally, we consider a multibaker chain joined together from two subchains 
with different material properties. 
This models a junction between two metals so that one can observe 
thermodynamic cross effects, like the Peltier and Seebeck effects, 
very much in the same arrangements as in classical experiments.

This paper is organized as follows. 
In Section II the local dynamics of the considered multibaker model
is defined, and its local entropy balance is worked out. 
In Section III, Model I -- Model IV are treated, which share periodic boundary 
conditions and represent thermodynamic settings of increasing complexity. 
Section IV is devoted to cross-effects. 
We conclude with a short discussion in Section V.

\begin{figure} 
{\hfill 
\epsfbox{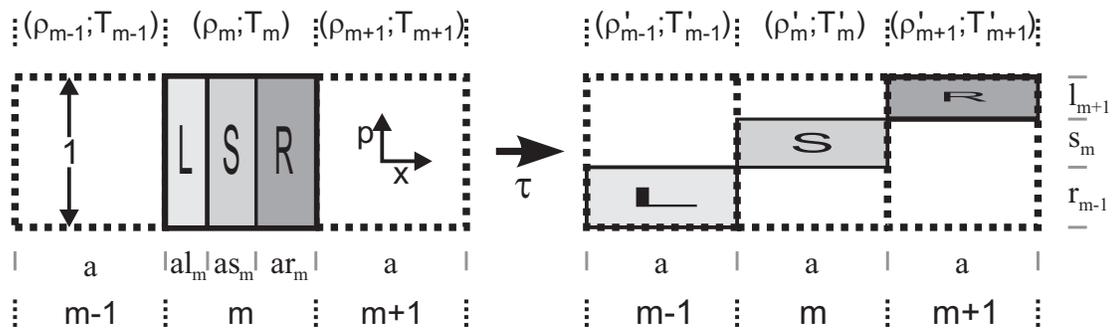} 
\hfill} 
\caption[]{
Graphical illustration of the action of the multibaker map on cell $m$. 
The letters $L$, $S$ and $R$ are inserted to visualize this action. 
Iteration of the rule after every time unit $\tau$ defines the time evolution. 
More details about the action of the mapping and the symbols needed for its definitions are 
given in the text. 
The symbols 
   $\varrho_i$ and $T_i$ 
indicated on the margins show the average values of the fields on the cells
and on its neighbors $i=m \pm 1$. 
\label{fig:iteration}}
\end{figure}
\section{Local transport and thermodynamic relations for multibakers}

In this section we describe the local dynamics of a cell of a multibaker map 
modelling a system 
with particle \cite{Gasp,VTB97,VTB98,BTV98,GD} and heat transport \cite{MTV99,MTV00}. 
We work out its density and kinetic-energy dynamics, and present general relations for the entropy 
changes.  
The effect of boundary conditions will be considered in subsequent sections for a few models with 
progressively richer thermodynamics. 

The phase space $(x,p)$ of the multibaker map consists of cells 
labelled by the index $m$ (Fig.~\ref{fig:iteration}). 
The division of the $x$ axis into cells corresponds to 
a partitioning of the configuration space into regions, sufficiently 
large to allow to characterize the state inside the cell by 
thermodynamic variables and small 
enough to neglect variation of these variables on the length scale of 
the cells {\em (local equilibrium approximation).\/} 
Every cell has a width $a$ and height $b \equiv 1$. 
The coordinates of individual particles in the cell are given as a position variable $x$, and a 
momentum-like variable $p$. 
We are interested in the dynamics of two dimensionless fields, the phase-space density $\varrho(x,p)$, 
and a field $T(x,p)$ characterizing the local kinetic energy per particle. 
After each time unit $\tau$, every cell is divided into three columns (Fig.~\ref{fig:iteration}) 
with respective widths 
   $a l_m$, $a s_m$ and $a r_m$. 
(Note that   
   $l_m + s_m + r_m = 1$ for any $m$.) 
The right (left) column of width
     $a r_{m}$ 
    ($a l_{m}$) 
is uniformly squeezed and stretched into a strip of width $a$ and of height
     ${l_{m+1}}$ 
    (${r_{m-1}}$),  
which is mapped to the right (left) neighbouring cell. 
The middle one preserves its area and remains in cell $m$.
Note that the map is one-to-one on its domain. 
It globally preserves the phase-space volume, 
but it can {\em nevertheless\/} locally expand
or contract the phase-space volume. 
In Ref.~\cite{VTB98,MTV00} it was argued that only the choice of 
the contraction factors given 
here can be consistent with thermodynamics 
(in fact, one can find an analogous formulation with a 
fully area preserving dynamics at the expense of a spatial variation of 
the volume of the cells of the multibaker; cf.~\cite{TG98}).

The field $T$ is advected by the particle dynamics, and --- in order to mimic a 
local heating of the system --- it is also multiplied by a factor $(1+ \tau q)$ 
depending on the averages 
characterizing the local currents and the thermodynamic state. 
By this a {\em mean-field-like\/} coupling of the motion of the particles in and 
around of a given cell is introduced. 
In general, the width of the columns may depend on the variables characterizing the 
thermodynamic state in the vicinity of the cell, so that they vary in time and space. 
This is indicated by the explicit dependence of the parameters 
on the cell index. 
Iteration of these rules defines the time evolution of the system. 
The $(x,p)$ dynamics generates ever refining 
structures in the distributions $\varrho(x,p)$ and $T(x,p)$. 
For simplicity, we take the fields initially constant in each cell:  
$\vrho(x,p) = \vrho_m $, $T(x,p) = T_m $ .

\subsection{Dynamics of the particle density and the particle current}
\label{sec:density}

After one step of iteration, the fields will be piecewise constant on the strips defined 
in Fig.~\ref{fig:iteration}. 
Due to the conservation of particles, the phase-space density takes the respective values 
\begin{equation}
   \varrho_{m,r}' =  \frac{r_{m-1}}{{l_{m}}} \; \varrho_{m-1} ,
\;\;\;\;
   \varrho_{m,s}' = \varrho_{m} ,
\;\;\;\;
   \varrho_{m,l}' =  \frac{l_{m+1}}{{r_{m}}} \; \varrho_{m+1}.  
\label{eq:rho_update}
\end{equation}
(The prime will always indicate quantities evaluated after one time step.) 
The contraction factors $r_{m-1}/l_{m}$ and $l_{m+1}/r_{m}$ contribute to  
the (weighted) local phase-space contraction $\sigma_m$ of cell $m$: 
\beq 
   \sigma_m 
= 
   \frac{1}{\tau} 
   \left[\vrho_{m-1} r_{m-1}  \ln\frac{r_{m-1}}{l_m} 
           + \vrho_{m+1} l_{m+1} \ln \frac{l_{m+1}}{r_m}
   \right]   .
\label{eq:contract}
\eeq    
After one time step, the average density $\vrho_m'$ in cell $m$ 
is determined by its initial density $\vrho_m$ and by the 
initial densities $\vrho_{m\pm 1}$ of the 
neighbouring cells. 
Multiplying the strip densities (\ref{eq:rho_update}) with the widths of the respective strips, 
adding them up and dividing the sum by the width $a$ of the cell, 
one obtains the average (or the {\em coarse-grained\/}) 
density after the iteration 
\beq
   \varrho'_m=s_m \varrho_m+ r_{m-1} \varrho_{m-1}+ l_{m+1} \varrho_{m+1}. 
\label{eq:update}
\eeq
The coarse-grained density evolves according to this 
master equation, which 
can be rearranged to obtain the discrete conservation law of the density 
\beq
   \frac{\vrho'_m-\vrho_m}{\tau}=-\frac{j_{m}-j_{m-1}}{a} . 
\eeq
Here 
\beq
   j_m=\frac{a}{\tau} (r_m \vrho_m - l_{m+1} \vrho_{m+1}) 
\label{eq:j}
\eeq
is the discrete particle current flowing through the right boundary of cell $m$.

\subsection{The kinetic-energy dynamics and the energy current} 

According to the $T$ dynamics described above, the updated values 
   $T_{m,r}'$, $T_{m,s}'$, $T_{m,l}'$ 
for $T$ on the respective strips $R$, $S$, $L$ of cell $m$ contain a source term 
characterized by a local strength $q_m$: 
\begin{eqnarray}
   T_{m,r}' = T_{m-1} \; \left[ 1 + \tau q_m \right] ,
\nonumber \\ 
   T_{m,s}' = T_{m}   \; \left[ 1 + \tau q_m \right] ,
\label{eq:T_update}
\\ 
   T_{m,l}' = T_{m+1} \; \left[ 1 + \tau q_m \right] .
\nonumber 
\end{eqnarray}
This strength is yet undetermined. 
It depends on the physical setting of thermostatting to be modelled and on the 
average of $\vrho$ and $T$ values in the cells and in its neighbours. 

The $(x,p)$ dynamics also drives the field $T$, i.e., 
after one iteration the kinetic-energy density of cell $m$ takes the value 
\beq 
   \varrho'_m T'_m 
=  
   [  s_m     \varrho_m     T_m 
    + r_{m-1} \varrho_{m-1} T_{m-1} 
    + l_{m+1} \varrho_{m+1} T_{m+1}] (1+\tau q_m)  .
\label{eq:wmprime0}
\eeq
This equation can be rearranged as a discrete balance equation for the time evolution of $\vrho T$: 
\begin{equation}
   \frac{\varrho_m^{'} T_m^{'} - \varrho_m T_m}{\tau} 
= 
    \varrho_m^{'} T_m^{'} \frac{q_m}{1 + \tau q_m} 
 -  \frac{j_m^{(\varrho T)} - j_{m-1}^{(\varrho T)}}{a} , 
\label{eq:wmprime1}
\end{equation}  
where 
$
   j_m^{(\varrho T)} 
= 
   T_m j_m  
 - ({a^2} \, l_{m+1}/{\tau})\; \varrho_{m+1}\, ({T_{m+1}-T_m})/{a} 
$ 
is a corresponding discrete energy current.
Note that the r.h.s of (\ref{eq:wmprime1}) is not a full divergence, in accordance with the fact that the kinetic energy is not a conserved quantity. 
In an isothermal system where there is 
no kinetic energy dynamics, no source can be present ($q_m = 0$).

\subsection{Gibbs entropy and the coarse-grained entropy}
\label{sec:GCE}

In this study we are interested in both the temporal evolution of 
the exact fields $\vrho (x,p)$ and $T(x,p)$, 
and in the evolution of their respective cell averages 
$\vrho_m$ and $T_m$. 
The former densities characterize the microscopic time evolution, 
while the averages describe the local thermodynamic state 
in spatially small regions. 
Both levels of description admit entropy functionals, 
which are commonly denoted as Gibbs and coarse-grained entropy. 

The Gibbs entropy $S^{(G)}$ is related to the detailed knowledge of the system. 
It is taken with respect to the exact densities $\vrho (x,p)$ and $T(x,p)$. 
In a given cell it is defined as 
\beq
   S^{(G)}_m
=
    - \int_{\hbox{over cell $m$}}  {\rmd} x \, {\rmd} p \; \vrho(x, p) \;
      \ln \left( \frac{\vrho(x, p)}{\vrho^{\star}} 
                                 T(x,p)^{-\gamma} \right) .
\label{eq:SG}
\eeq 
Here $\vrho^* T^{\gamma}$ plays the role of a local $T$-dependent reference density with a constant reference density ${\vrho^{\star}}$ and $\gamma$ an as yet undetermined constant. 

The coarse-grained entropy $S_m$ has a similar form, 
but it is based on the averaged values in the considered cell: 
\beq
   S_m = - a \vrho_m 
         \ln\left( \frac{\vrho_m}{\vrho^{\star}} T_m^{-\gamma} \right).
\label{eq:Sm} 
\eeq

As mentioned above, throughout the paper we only consider initial distributions, 
which are uniform in every cell (cf.~\cite{VTB98,GD,MTV00} for more general choices). 
As a consequence, initially $S_m =S^{(G)}_m$, and after one time step 
the entropies become 
\beq 
   S_{m}^{(G)'}
=
- a \; \bigg[
      s_m \varrho_m
      \ln\left( \frac{\varrho_m}{\vrho^{\star}}\; T_{m,s}^{'-\gamma} \right)
   + r_{m-1} \varrho_{m-1}
      \ln\left( \;
                \frac{\varrho_{m,r}'}{\vrho^{\star}} \; T_{m,r}^{'-\gamma} 
                \right)
  + l_{m+1} \varrho_{m+1}
      \ln\left( \;
                \frac{\varrho_{m,l}' }{\vrho^{\star}} \; T_{m,l}^{'-\gamma}
                \right)
\bigg] ,
\label{eq:SGprime} 
\eeq
and 
\beq
   S'_m = - a \vrho'_m 
         \ln\left( \frac{\vrho'_m}{\vrho^{\star}} T_m^{'-\gamma} \right).
\label{eq:Smprime} 
\eeq

\subsection{Entropy balance}
\label{sec:balance}

The coarse-grained entropy fulfills a local entropy balance in direct analogy to the one in irreversible thermodynamics. 
To derive this equation one identifies at any given time the difference $S_m-S^{(G)}_m$ as the information on the microscopic state of the system, which cannot be resolved in the coarse-grained description. 
The temporal change of this lack of information is then identified with the irreversible entropy production $\Delta_i S_m$, and the change $(S'^{(G)}_m-S^{(G)}_m)$ of the Gibbs entropy with the entropy flux $\Delta_e S_m $.          
Thus, 
\beq
\frac{S'_m-S_m}{\tau} = \frac{\Delta_e S_m}{\tau} +\frac{\Delta_i S_m}{\tau}  
\eeq
which is a discrete analog of (\ref{eq:entrbg}).

The form of the entropy production is [cf.~(\ref{eq:SGprime}) and 
(\ref{eq:Smprime})]: 
\begin{eqnarray}
   \frac{\Delta_i S_m}{\tau} 
& = & 
  \frac{[S'_m-S^{(G)'}_m] - [S_m-S^{(G)}_m]}{\tau} 
\non\\
&= &  
 \frac{a}{\tau} \; \bigg[  
   - \varrho_m' \ln\left( 
   \frac{\varrho_m' T_m'^{-\gamma}}{\varrho_m T_{m,s}'^{-\gamma}} 
              \right)
   + \varrho_{m-1} r_{m-1} 
      \ln\left( 
         \frac{\varrho_{m,r}' T_{m,r}'^{-\gamma}}{\varrho_m T_{m,s}'^{-\gamma}}
      \right) 
   + \varrho_{m+1} l_{m+1} 
      \ln\left( 
         \frac{\varrho_{m,l}'T_{m,l}'^{-\gamma}}{\varrho_m T_{m,s}'^{-\gamma}}
         \right)    
\bigg] , 
\label{eq:DiSm} 
\end{eqnarray} 
where we used that $S_m-S^{(G)}_m$ vanishes due to 
the particular choice of initial conditions. 

The entropy flux becomes 
\begin{equation} 
   \frac{\Delta_e S_m}{\tau} 
 =   - \frac{a}{\tau} 
\left[ 
   (\varrho_m^{'} - \varrho_m) \; 
   \ln\left( \frac{{\varrho_m}}{\vrho^*} T_m^{-\gamma} \right) 
+
   \varrho_m^{'} \ln \frac{T_{m,s}'^{-\gamma}}{T_m^{-\gamma}}
+
   \varrho_{m-1} r_{m-1} 
      \ln\left( \frac{\varrho_{m,r}'}{\varrho_m} 
                \frac{T_{m,r}'^{-\gamma}}{T_{m,s}'^{-\gamma}} 
         \right) 
+ 
   \varrho_{m+1} l_{m+1} 
      \ln\left( \frac{\varrho_{m,l}'}{\varrho_m}
                \frac{{T_{m,l}'}^{-\gamma}}{{T_{m,s}'}^{-\gamma}} 
         \right) 
\right]
\label{eq:DeSm} 
\eeq 
which can be split into a divergence of an entropy current 
and a flux into the thermostat
\beq   \label{eq:DeSmsplit} 
\frac{\Delta_e S_m}{a\tau} = -\frac{j_m^{(s)}- j_{m-1}^{(s)}}{a} 
                             +\Phi^{\rm (th)}_m 
\eeq
with  
\begin{mathletters}
\bega    
j_m^{(s)}
& \equiv &
 -  j_m  \ln\left( \frac{\varrho_m}{\varrho^{\star}} T_m^{-\gamma} \right)
 +       \frac{a l_{m+1}}{\tau} \varrho_{m+1}
      \ln  \left( \frac{\varrho_{m+1}}{\varrho_m} 
                  \frac{T_{m+1}^{-\gamma}}{T_m^{-\gamma}}      \right)
  - \vrho_{m} v_{m} , 
           \label{eq:js-mikro}          \\  
\Phi^{\rm (th)}_m 
& \equiv &
- \frac{1}{\tau} \left[
    \varrho'_m \ln\frac{T_{m,s}^{'-\gamma}}{T_m^{-\gamma}}
    + r_{m-1} \varrho_{m-1} \ln \left(
                               \frac{ r_{m-1} }{l_m}
                               \frac{T_{m,r}^{'-\gamma} T_m^{-\gamma}}
                            {T_{m,s}^{'-\gamma}  T_{m-1}^{-\gamma} }     \right)   
   +
      l_{m+1} \varrho_{m+1} \ln \left( 
                               \frac{{l_{m+1} }} {r_m}
                               \frac{ T_{m,l}^{'-\gamma} T_m^{-\gamma} }
                            { T_{m,s}^{'-\gamma}  T_{m+1}^{-\gamma} }       \right)
\right]      \non\\
&-&
\frac{v_m \vrho_m - v_{m-1} \vrho_{m-1}}{a} .
\label{eq:phith}
\ega   
\label{eq:jmsPhimth}
\end{mathletters}
Note that (\ref{eq:DeSmsplit}) is a discrete counterpart of (\ref{eq:balance0}), 
and 
        $j_m^{(s)}$ and $\Phi_m^{(th)}$ 
are the discrete entropy current and entropy flux to the thermostat, 
respectively.

\subsection{The macroscopic limit}

The projection of the multibaker dynamics on the $x$ axis
corresponds to a biased random walk with some diffusion coefficient
and drift. 
The drift has to be present if we want to model nonequilibrium systems 
subjected to electric fields and/or temperature gradients. 
The requirement of consistency with an advection diffusion equation
in the large system and long time limit, when
the cell size is  much smaller than the system size, and
the time unit is much shorter than the macroscopic relaxation time,  
leads \cite{VTB97,VTB98,BTV98,MTV99} to the scaling relation:  
\begin{mathletters}
\bega  
   r_m &=& \frac{\tau D}{a^2} \left( 1+\frac{av_m}{2D} \right) ,
\label{eq:rm}
\\
   l_m &=& \frac{\tau D}{a^2}\left(1-\frac{av_m}{2D} \right) .
\label{eq:lm}
\ega   \label{eq:rmlm}
\end{mathletters}
Here we allow for a location dependence of the drift $v_m$ but assume the diffusion coefficient $D$ 
to be spatially constant. 
The continuum limit of the multibaker dynamics, 
which is taken with these constraints, is called  
the {\em macroscopic limit.\/}  
Formally, it corresponds to taking $a, \tau \rightarrow 0$ 
while keeping $D$ fixed, and $am$, 
$\vrho_m$, $T_m$, $v_m$ and $q_m$ finite so that they approach a macroscopic position coordinate $x$, 
and smooth functions $\vrho (x)$, $T(x)$, $v(x)$ and $q(x)$, respectively. 
After taking this limit we call $\vrho (x)$ the density and $T(x)$ temperature distribution in 
the system. 
The macroscopic limit of all the local relations given in 
(\ref{eq:update})--(\ref{eq:DeSm}) can 
be worked out explicitly \cite{VTB97,VTB98,BTV98,MTV99}. 
In the following this limit will be indicated by an arrow `$\rightarrow$' . 
Here we only mention that the system's equation of state 
turns out \cite{MTV99,MTV00} to be that of classical ideal gas 
with $\gamma\vrho$ as its heat capacity (measured in units of 
Boltzmann's constant).

\begin{figure} 
{\hfill
\epsfbox{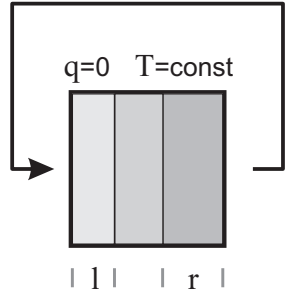} 
\hfill} 
\caption[]{Isothermal single-cell periodic baker map. The right boundary is
identified with the left one.
\label{fig:model1}}
\end{figure}
\section{Periodic models}

\subsection{Model I: Isothermal single-cell multibaker}

We start by discussing the simplest conceivable model for describing a macroscopic transport 
process. 
A particle current induced by an external field in an isothermal environment described  
by a single baker cell subjected to periodic boundary conditions. 
The right boundary of the cell is identified with its left boundary and the mapping is from 
the cell onto itself. 
Because of driving $r \not = l$, and thermostatting is applied 
via the appearance of the contraction rates $l/r$ and $r/l$ 
in order to reach a steady state. 
This mapping propagates the coordinates 
of a large number of particles, which do not interact, i.e., they all are 
mapped by the same mapping. 
Clearly, this system does not admit a spatial resolution of the densities characterizing the 
transport process. 
Its local and global behaviour coincides, so that the subscript $m$  of the densities can be 
descarded in this case. 
 
For the single-cell multibaker the master equation (\ref{eq:update}) predicts 
$ 
   \varrho ' = \varrho \equiv \bar\vrho .
$  
This implies that the model is describes tranport in a steady state with the average density
$\bar\vrho$. 
The particle curent 
$
   j = (a/\tau) (r-l)\bar\vrho = v \, \bar\vrho 
$
is constant in space and time, and the entropy production 
(\ref{eq:DiSm}) becomes ($T=const.$).
\beq
\frac{\Delta_i S}{a\tau} = 
\bar\vrho \frac{(r-l)}{\tau}  
\ln \left( \frac{r}{l} \right) \equiv \sigma 
\eeq
which has the macroscopic limit:
\beq
\frac{\Delta_i S}{a\tau}\rightarrow \sigma^{(irr)} = \bar\vrho \frac{v^2}{D}.
\eeq
In a similar way, the entropy flux 
(\ref{eq:DeSmsplit}, \ref{eq:phith}) 
has the macroscopic form: 
\beq  \label{eq:DeS1}
\frac{\Delta_e S}{a\tau}\rightarrow \Phi^{\rm (th)} = - \bar\vrho \frac{v^2}{D}
\eeq
As expected in a steady state, $\Delta_i S$ and $\Delta_e S$ add up to zero. 
More interestingly, however, these contributions to the change of entropy are also 
directly proportional to the local phase-space contraction \eq{contract}, which 
reduces to $\bar\vrho \, v^2/D$ in the macroscopic limit. 

\begin{figure} 
{\hfill
\epsfbox{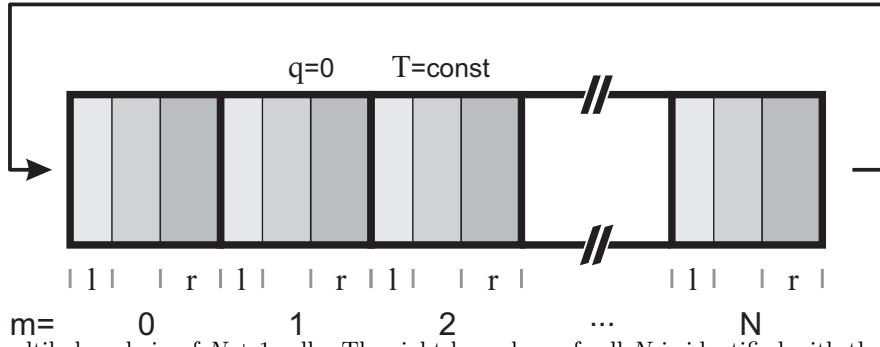} 
\hfill}
\caption[]{Isothermal multibaker chain of $N+1$ cells. The right
boundary of cell $N$ is identified with the left boundary of
cell $0$.
\label{fig:model2}}
\end{figure}
\subsection{Model II: Isothermal multibaker chain for field driven transport}

In the single-cell multibaker the coarse-grained density cannot evolve in time, 
since no spatial variations are resolved. 
For this reason, one also cannot distuinguish between the local entropy balance
as described by \eq{cs}, \eq{balance0} and the balance 
\eq{entrbg} for the full macroscopic system. 
In order to address these points, we generalize the previous setting by considering a 
multibaker chain of $N+1$ cells, with spatially constant driving 
(i.e., $r_m =r $ and $l_m=l $) and periodic boundary conditions: cell $N+1$ and cell $0$ are
identified.  
{F}rom (\ref{eq:j}) we obtain for the particle current 
\beq
   j_m = \frac{a}{\tau} \left[ (r-l) \vrho_m - l (\vrho_{m+1}-\vrho_m) \right]
\label{eq:j2}
\eeq 
which has the macroscopic limit:
\beq \label{eq:j20}
   j_m \rightarrow j\equiv \vrho v-D\partial_x \vrho.
\eeq
This current varies along the chain as long as the cell densitites evolve in time. 
The asymptotic state, however, is formed by a spatially uniform density distribution
with the average density $\bar\vrho$.

For constant temperature $T$ the irreversible entropy production (\ref{eq:DiSm}) becomes 
\beq   \label{eq:DiSmchain1}
\frac{\Delta_i S_m}{a\tau} =
\frac{1}{\tau} 
\bigg[
-\vrho'_m 
 \ln \frac{\vrho'_m}{\vrho_m} 
 + r \vrho_{m-1}\ln \left( \frac{\vrho_{m-1}}{\vrho_{m}} \frac{r}{l} \right) 
+ l \vrho_{m+1}
 \ln \left( \frac{\vrho_{m+1}}{\vrho_m} 
 \frac{l}{r} \right) 
\bigg] .
\label{eq:DiSm1}
\eeq

{ Using (\ref{eq:update}) and (\ref{eq:rmlm}), a lengthy but 
straightforward calculation for $a, \tau \rightarrow 0$ 
\cite{MTV00} shows that} 
this {\it local} form is consistent with thermodynamics 
since in a general non-steady state it approaches 
\beq   \label{eq:sigmairr2}
\sigma^{(irr)} = 
\frac{{(\vrho v-D\partial_x \vrho)}^2}{\vrho D}
= \frac{j^2}{\vrho D} 
\eeq
in the macroscopic limit. 
In a similar way \cite{VTB98,BTV98}, the entropy flux 
\eq{DeSm} can be calculated, 
which in the macroscopic limit has the form (\ref{eq:balance0}),   
with the entropy current  
\beq
   j^{(s)} = - j\; \left[1+ \ln (\varrho/\varrho^*) \right]
\eeq 
and the entropy flux 
\beq  
   \Phi^{(th)}=-\frac{vj}{D}
\label{eq:phith2}
\eeq
transfered directly to the environment. 
This local expression expresses that every cell is coupled to the thermostat. 

Due to the additional spatial resolution (as compared to Model I) the 
local and global features of the entropy balance can be different. 
It is worth considering the global entropy production, defined as 
[cf.~(\ref{eq:DiSmchain1})] 
\beq
\sum_{m=0}^{N} \frac{\Delta_i S_m}{\tau}
=
\frac{a}{\tau}
\left[ \sum_{m=0}^{N} \vrho_m \right] (r-l) \ln \left( \frac{r}{l} \right) 
+
\frac{a}{\tau}
\sum_{m=0}^{N} [-\vrho'_m \ln \vrho'_m + \vrho_m \ln \vrho_m ]  .    
\eeq
{ Here, the logarithm of the ratios of densities of neighbouring cells 
in (\ref{eq:DiSm1}) drop out in the sum due to the periodic boundary 
conditions. The global entropy production then takes} 
the macroscopic form 
\beq
\sum_{m=0}^{N} \frac{\Delta_i S_m}{\tau}
\rightarrow
\frac{d_i S}{dt}
= 
\left[ \sum_{m=0}^{N} a \vrho_m \right] 
\frac{v^2}{D}-\sum_{m=0}^{N} a \partial_t [\vrho \ln \vrho]  
= 
{\cal N} \frac{v^2}{D} 
+ \frac{d S}{dt} , 
\eeq
where ${\cal N} = a\bar\vrho (N+1)$ is the total number of particles 
in the multibaker chain, and 
$
   S=\sum_m S_m  
$
is the total entropy. 

The global form of the entropy flux is [cf. \eq{DeSm}] 
\beq
   \sum_{m=0}^{N} \frac{\Delta_e S_m}{\tau}
=
-\frac{a}{\tau} 
    \left[ \sum_{m=0}^{N} \varrho_m \right] (r-l) 
                   \ln \left( \frac{r}{l} \right) 
= - \sum_{m=0}^N  a \sigma_m ,
\eeq
which takes the macroscopic limit 
\beq  \label{eq:deSdt}
   \sum_m \frac{\Delta_e S_m}{\tau} 
\rightarrow 
  \frac{d_e S}{dt}=
- {\cal N} \frac{v^2}{D} .   
\eeq
This shows that the local and the global entropy balances are 
markedly different. 
Locally, the entropy current depends only on 
the local current and the local density. 
There is only an indirect influence of the drift velocity $v$ 
through its contribution 
to the current and its influence on the density profile. 
In contrast, the macroscopic flux only depends on the total number of particles in the system 
and on the drift velocity. 
It is constant in time since it neither depends on the 
current nor on the density profile, which in general both evolve in time. 
Interestingly, also in this more general setting the negative of the global entropy flux  
equals the (total) phase space contraction at any time. 
{ This is in full harmony with the result obtained for 
the entropy flux in noisy dynamical systems by Nicolis and Daems 
\cite{ND,DN}. } 
In contrast, the total irreversible entropy production is in general no longer 
directly related to the phase-space contraction. 
Its local form exactly amounts to Joule's heating $j^2/\varrho D$, and globally 
it picks up an additional contribution characterizing the time-evolution of 
the macroscopic states. 
Therefore, the often cited relation between the global entropy production and 
the phase-space contraction only holds in steady states, where the 
modulus of the entropy flux coincides with the rate of entropy
production.

\begin{figure} 
{\hfill
\epsfbox{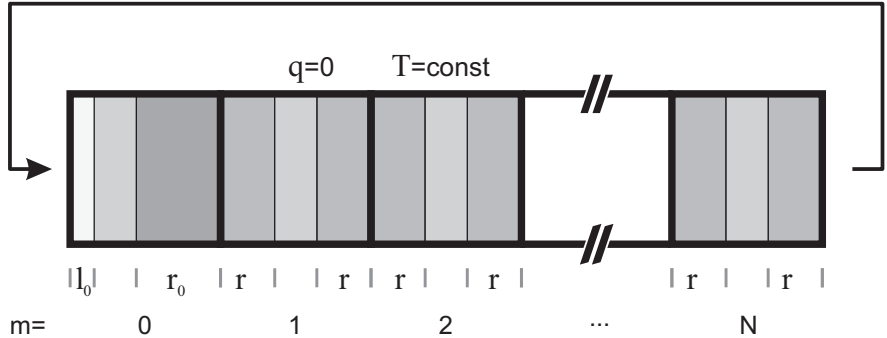} 
\hfill}
\caption[]{Isothermal multibaker chain with driving in cell $0$.
There is no
drift in the bulk.
\label{fig:model3}}
\end{figure}
\subsection{Model III: Isothermal multibaker chain with driving localized 
to a single cell}

A generalization of the previous case is a system modelling boundary-driven transport. 
Here, we consider a slightly different setting with periodic boundary conditions. 
Away from the point of driving the macroscopic properties correspond to  
that of boundary driven transport, but due to the periodic boundary conditions 
the model has a simpler structure as a dynamical system, which makes the analogy 
with our previous models more transparent. 
 
We consider a multibaker chain where 
cell $0$ has a behaviour different from the rest of the chain 
in the sense that its drift velocity is different from that of the bulk 
where $r=l$,  
(i.e., $v=0 $) but the diffusion coefficient is the same. 
This implies 
$ r_0 - r \neq l - l_0 \not= 0 $ 
(cf. (\ref{eq:rmlm})).  
The driving in cell zero has the tendeny to generate an accumulation 
of particles right to it and a slowly decreasing density distribution 
on its left (we assume $v_0 >0$, $r_0 >r$). 
In a steady state, this leads to a linear density profile 
\beq
\vrho_m 
= 
\bar\vrho + \left( \frac{N-1}{2} -m +1 \right) \delta \vrho 
\quad\hbox{for $m=1...N$,}
\label{eq:StStat} 
\eeq
whose increment $\delta \vrho$ is uniquely determined by $r_0$ (or $v_0$). 
In cell zero we find 
   $\vrho_0=\bar\vrho$. 
A substitution into the master equation for 
$m=1$ or $N$ leads to 
\beq
   \delta \vrho = \bar\vrho \frac{2}{N+1} (\frac{r_0}{r}-1) .
\eeq
Again the average density $\bar\vrho$ is related to the number of particles 
${\cal N}$ in the system via $a\bar\vrho (N+1) = {\cal N}$. 
By taking into account that according to 
(\ref{eq:rmlm}) 
$r_0 = r [1+av_0 / (2D)]$,
we obtain that  $\delta \vrho$ is indeed proportional to $v_0$: 
\beq  \label{eq:v0}
v_0 = D \frac{\delta \vrho}{a} \frac{N+1}{\bar\vrho} 
    = D \frac{\delta \vrho}{a} \frac{{\cal N}}{a {\bar\vrho}^2}. 
\eeq
The steady state current is: 
\beq
j= \bar\vrho \frac{v_0}{2}
- D \frac{\vrho_1 - \bar\vrho}{a} 
 =  D\frac{\delta \vrho}{a}
 =\frac{v_0 \bar\varrho}{N+1}. 
\eeq
The diffusion current of cells $N$ and $0$ is much stronger 
than otherwise, due to the sudden jump in the densities, but in the steady state
the surplus is exactly compensated by the local drift currents. 
Thus the particle transport in the steady states of Model II 
and Model III are equivalent provided the drift $v$ in the former 
coincides with $ D \delta\vrho / (a \bar\vrho)$ in the latter. 
As far as the steady state transport is concerned, it does not matter, 
whether one applies a 
small uniform field leading to a spatially uniform drift $v$ or a large one 
$v_0 = (N+1)v$ in a single cell only. 

All thermodynamic relations of relevance can be worked out not only for
the steady state (\ref{eq:StStat}), but for general non-steady states. 
The local forms of the entropy production and the entropy flux in the bulk  
are special cases of (\ref{eq:sigmairr2})-(\ref{eq:phith2}). 
Since $r=l$ in the bulk, there is no entropy flux 
flowing into the thermostat $\Phi^{(th)}=0$ for $m=2,...N-1$.  
On the other hand, the entropy flux in cell $0$ 
and its neighbours does contain a part which cannot be written as a 
divergence. 
Due to (\ref{eq:phith}), the average density of the flux flowing into the 
thermostat turns out to be:
\bega
\Phi^{\rm (th)}
\equiv
\Phi_N^{\rm (th)}+\Phi_0^{\rm (th)}+\Phi_1^{\rm (th)} 
&=&
\frac{1}{ \tau} 
\bigg[  (r \vrho_1 - r_0 \vrho_0)  \ln \left( \frac{r_0}{r} \right) 
     -(r \vrho_N - l_0 \vrho_0)  \ln \left( \frac{r}{l_0} \right) \bigg] \non\\ 
&\approx &  
 \frac{v_0}{2} \frac{\vrho_1-\vrho_N}{a} 
- \vrho_0 \frac{v_0^2}{4D} 
- \frac{\vrho_1+\vrho_N}{2} \frac{v_0^2}{4D} .
\ega
In the last approximation we have assumed that $a v_0 \ll 2 D$.

In the steady state $\vrho_N - \vrho_1 = - (N-1) \delta \vrho = - (N-1)aj/D$, 
$\vrho_0 = \frac{(\vrho_1+\vrho_N)}{2}=\bar\vrho $,  
and thus 
\beq
\Phi^{(th)}
\equiv 
-\frac{v_0 j}{D} .
\eeq 
In the macroscopic limit the quantity 
$\delta \vrho /a$ approaches the gradient $-\partial_x \vrho$ in the 
bulk, $\bar\vrho$ and ${\cal N}$ stay constant and thus 
$v_0$ in \eq{v0} is proportional to $1/a$. The driving is singularly strong and 
so is the entropy flux density into the thermostat. 
By integrating, however, over the volume of cell zero and its neighbours we obtain 
the total entropy flux into the thermostat 
$-a v_0 j/D = -j^2 {\cal N} /{\bar\vrho }^2 D $. 
It coincides with the macroscopic limit of the total entropy flux 
   $ \sum_{m=0}^N \Delta_e S_m /\tau $ 
since the integral of $\partial_x j^{(s)}$ vanishes in a periodic system, i.e., 
\beq
   \frac{d_e S}{dt} = -\frac{j^2}{\bar\vrho D} \frac{{\cal N}}{\bar\vrho}
\eeq
This result is equivalent to the steady state version of (\ref{eq:deSdt}) expressed by 
the current. 

In this model there is no need for taking out heat along the bulk,
thermostatting is active in cell $0$
and its neighbours only. 
It extracts exactly the same entropy flux there as the full entropy flux of Model~II 
in the steady state \cite{CL}.
Thus the models refer to two different realizations of thermostatting the 
transport process. 
Model II should be viewed as e.g. a wire which is kept at constant temperature 
by removing the heat due to dissipation everywhere along its length --- 
Model III is closer related to a thermally isolated system, where heat 
is tranported to the ``boundary'', from where the system is driven.  
For the multibaker this takes place in the special cell $m=0$. 
Boundary driven transport typically leads, however, 
to non-uniform temperature profiles. 
A full treatment of such transport processes should be based therefore on 
a multibaker chain with kinetic energy dynamics. 

\begin{figure} 
{\hfill
\epsfbox{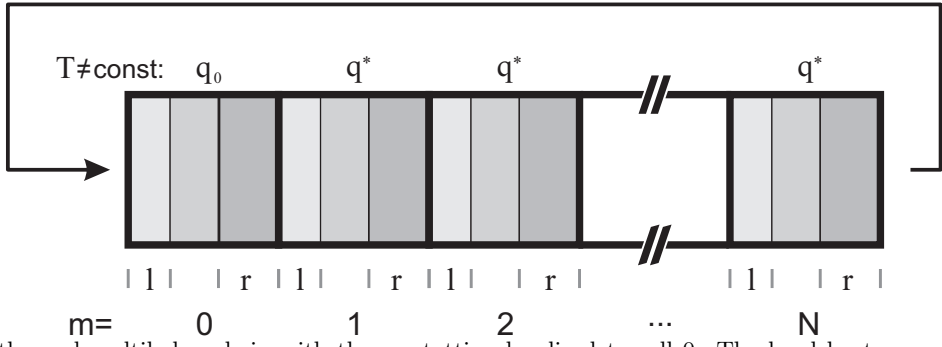} 
\hfill}
\caption[]{
Non-isothermal multibaker chain with thermostatting localized to cell $0$.
The local heat source $q_0$ of that cell is different from that in the bulk, 
which is 
   $q^*=-vj/D$.
\label{fig:model4}}
\end{figure}
\subsection{Model IV: Multibaker chain with thermostatting localized to a single cell}

In the realm of classical thermodynamics 
heat is transported to the boundaries of the system. 
If, however, there is a break in the temperature profile, a 
jump in the heat current occurs and heat is taken out at this  
point. 

In order to model such a situation, we consider 
a multibaker chain with kinetic-energy dynamics. 
The general relation 
(\ref{eq:DiSm}) 
{ and a calculation similar to the one leading to (\ref{eq:sigmairr2})} 
yields in the macroscopic limit for the entropy production \cite{MTV00} 
\beq
\sigma^{(irr)}=  \lambda
                  {\left( \frac{\ptx T}{T} \right)}^2 
                  +\frac{j^2}{\vrho D}
\eeq 
where 
$j=\vrho v -D \partial_x \vrho$ is the particle current, and 
$\lambda=\gamma \varrho D$ is the heat conductivity of the model.
{F}rom  \eq{DeSm}-\eq{phith} we obtain  
\beq   \label{eq:phith4}
   \Phi^{(th)} = \gamma\vrho  q - \frac{vj}{D} 
\eeq
as the entropy flux let directly into the thermostat, and 
\beq  \label{eq:js}
   j^{(s)}= -\lambda \frac{\ptx T}{T} + \frac{e\Pi}{T} j , 
\eeq
as the entropy current with $\Pi$ the bulk Peltier coefficient 
\beq
\frac{e\Pi}{T} = - \left( 1 +  \ln \frac{\vrho T^{-\gamma}}{\vrho^*} \right) .
\label{eq:Pi}
\eeq 
Note that there is always a possibility to 'close' the system locally in the 
sense that the source term 
$q=q^*=vj/\lambda$ is chosen such that $\Phi^{(th)}$ vanishes.

We consider a periodic chain 
with  fixed transition probabilities ($r_m=r$, $l_m=l$, $m=0.1,..,N$).
The local heating sources are assumed to be constant in the bulk:
$q_m=q^*$ for $m=1,...,N$ which differs from the source $q_0$ of cell
$0$.
In the steady state we find then 
a constant particle density along the chain.  
Inside the bulk, the kinetic-energy equation (\ref{eq:wmprime0}) implies  
for the steady temperature distribution 
\beq
   T_m = [ (1-r-l) T_m +r T_{m-1}+ l T_{m+1}](1+ \tau q^* ) .     
\label{eq:Tmstat}
\eeq 
With periodic boundary conitions 
($ T_{m=0}=T_0 $, and $T_{m=N+1}=T_0$)
this equation has the following 
general solution: 
\beq \label{eq:Tm}
T_m = \frac{T_0}{\sin[b(N+1)]} 
      { \bigg( \frac{r}{l} \bigg) }^{\frac{m}{2}}  
      \left\{ \sin[b(N+1)-bm] + 
     { \left( \frac{l}{r} \right) }^{\frac{N+1}{2}} 
       \sin(bm) \right\}
\eeq
where 
\beq
\cos b 
= 
\sqrt{rl} \left( 1-\frac{\tau q^\star}{(1+\tau q^\star) (r+l)}  \right) .
\eeq
The solution (\ref{eq:Tm}) has a break in cell zero,  
in the sense that the left and right derivatives are different.
Only at the end of the system is the entropy flux not a full divergence.  
Applying the kinetic-energy equation \eq{wmprime0} to cell zero  
in a steady state, where the density is constant, we find that: 
\beq  \label{eq:q0}
   q_0 
=
   \frac{1}{\tau} \frac{r(T_0-T_N)+l(T_0-T_1)}{(1-r-l) T_0+r T_N +l T_1} .
\eeq
In the macroscopic limit 
\beq
   a q_0 
\rightarrow 
   \frac{D}{T}
   \left[ \partial_x \left. T \right|_{(-0)} -\partial_x \left. T 
                     \right|_{(+0)} \right]  . 
\eeq  
This implies that the source density $q_0$ is singular 
but the  total source $Q_0=a q_0$ inside cell $0$, 
is finite.


It is worth comparing this with the thermodynamic treatment of the same problem. 
If there is a jump in the entropy current,  
in order to have finite entropy flux density $\Phi$ in each point,  
it is unavoidable to allow for a 
$\tilde{\Phi}$ which is not a full divergence: 
\beq  \label{eq:flux}
\Phi = - \partial_x j^{(s)} + \tilde{\Phi} .
\eeq
The form of $\tilde{\Phi}$ one obtains by integrating (\ref{eq:flux}) 
around the point where the jump in the derivative appears ($x=0$): 
\beq
\int_{-\epsilon}^{\epsilon} \Phi dx = 
           - \left. j^{(s)} \right|^{(+\epsilon)}_{(-\epsilon)} + 
                             \int_{-\epsilon}^{\epsilon}  \tilde{\Phi} dx  . 
\eeq
The smoothness of 
$\Phi$ implies that for 
$\epsilon\rightarrow 0$  
\beq   \label{eq:tildephi}
\int_{-\epsilon}^{\epsilon} \tilde{\Phi} dx = 
                    - \left. j^{(s)} \right|^{(+0)}_{(-0)}.
\eeq
The multibaker result \eq{phith4} implies that if 
$q$ is singular as in cell 0, then 
$\Phi_0^{(th)}=\gamma \vrho q_0$. We then immediately see 
that $\Phi_0^{(th)} $ is the analog of $\tilde{\Phi}$.
Indeed, by inserting the expression \eq{js} for 
$j^{(s)}$ we have: 
\beq
\int_{-\epsilon}^{\epsilon} \tilde\Phi dx = 
\frac{\lambda}{T}
[\partial_x \left. T \right|_{(+0)} -\partial_x \left. T \right|_{(-0)}]
\eeq
which, on account of  $\lambda = \gamma \vrho D$,    
exactly corresponds to (\ref{eq:q0}). 

We have shown, that the thermodynamic evaluation and the macroscopic limit 
of $q_0$ lead to the same result.
Physically this means, that by a proper choice of the source terms even 
the singularity in the entropy flux can be described in full harmony 
with thermodynamics, and the flux let to flow in the thermostat 
is exactly the amount of heat what is taken out also 
in the thermodynamic description if a break appears.

\begin{figure} 
\epsfbox{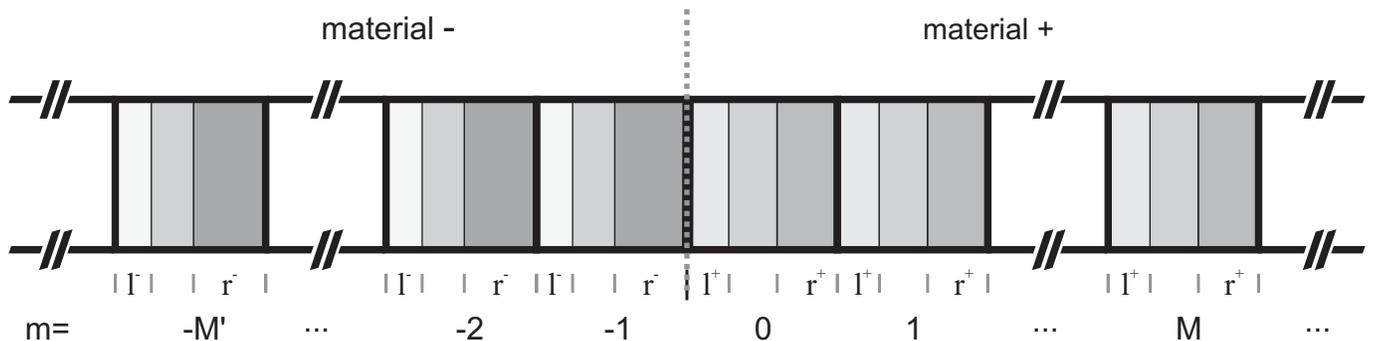} 
\caption[]{Two long multibaker chains,  
representing materials $-$ and $+$, joined together at the junction
between cells $-1$ and $0$. The leads are in cells $-M'$ and $M$.
\label{fig:model5}}
\end{figure}
\section{Cross effects in a multibaker model} 

Thermodynamic cross effects, which probe the (off-diagonal) Onsager coefficients, 
are difficult to observe in homogenous systems. 
When two materials are put into contact, however, 
they play a dominant role in understanding the heat and entropy currents. 
In order to mimick such phenomena, we consider two multibaker chains 
containing the cells $m = -M' \cdots -1$ and $m = 0 \cdots M$, respectively, 
$M$, $M'>>1$ 
which are brought into contact at $m=0$ (cf.~Fig.~\ref{fig:model5}). 
Now, the parameters 
$l^-$, $s^-$, $r^-$ and $l^+$, $s^+$, $r^+$ 
in the two parts are different, and for generality we will also assume that 
the constant reference densities $\varrho^{\star, \pm}$ are different. 
These differences will represent the different thermodynamic and transport 
properties of the materials. 
The difference in $r$ and $l$ gives rise to different 
drifts (conductivities) and 
diffusion coefficients, and the one in the reference density might be thought 
of reflecting for instance a different effective mass of the electrons.

As in the previous subsection, the dynamics of this multibaker chain drives a density 
and a kinetic-energy field. 
In order to simplify the structure of the steady state density profiles, we
restrict to the case 
        $r^+ / l^+ = r^- / l^-$. 
This choice is motivated by a physical interpretation of $r/l$. 
After all, the macroscopic limit of $r/l$ is $1+av/D$, 
and $v/D$ is 
proportional to the external (electric) field, such that the requirement 
expresses that the external field should be the same in both materials. 
In the remainder of this section we discuss the transport in this model 
in two different settings: 
(i) a constant (non-vanishing) particle current and constant temperature; 
(ii) vanishing particle current and an isolated system which is only 
thermostatted 
at the ``junction'' $m=0$ and at the two ``leads'' $m=-M'$ and $m=M$, respectively. 
Setting (i) allows us to discuss the Peltier effect and setting (ii) 
is used for the Seebeck effect. 

Before turning to these specific settings, however, we discuss the 
steady state 
profile of the (particle) density in general  
One does not expect noticable gradients in the electron density in 
either material, so that we fix them to the constant values 
        $\varrho^-$ and  $\varrho^+$, 
leading to the spatially uniform current (cf. \eq{j}) 
\[ 
        j = \frac{a}{\tau} (r^+ - l^+) \, \varrho^+ 
          = \frac{a}{\tau} (r^- - l^-) \, \varrho^- . 
\] 
In order to have the same current also across the junction, 
one has to require 
   $l^- \varrho^- = l^+ \varrho^+$ in adition. 
Together with the fact that $v/D$ is fixed for the whole system, this implies 
that there is  a constant amount of Joule's heating $vj/D$ per unit length of the 
system, which either has to be transferred to a local thermostat  
(cf. \eq{phith4}), or leads to a 
local heating, i.e., enforces non-uniform temperature profiles.

\subsection{The Peltier effect}

The requirement of a constant temperature in the setting of the Peltier effect 
requires the use of a thermostatted dynamics, $q=0$. 
In that case the Joule heating is transferred to the thermostat. 
Away from the junction, this leads to the flux $\Phi^{(th)} = vj/D$.
In the entropy balance the difference in the materials 
shows up {\em only\/} in the entropy currents. 
In view of (\ref{eq:js}), they become different in the two parts of the multibaker 
\[ 
        j^{(s, \pm)} = - j \; \bigg( 1+\ln\left[ 
           \frac{\varrho^\pm}{\varrho^{\star\pm}} \, T^{-\gamma} \right]  \bigg) 
           = \frac{e \Pi^\pm}{T} j 
\] 
implying that at the junction (i.e., between cells $m=-1$ and $m=0$) 
an additional 
heat flux, the {\em Peltier heat\/}, is directed to the thermostat. 
It is characterized by the difference of the entropy currents 
\[ 
        j^{(s, +)} -    j^{(s, -)}      
= 
        -j \; 
        \ln\frac{\varrho^+ /\varrho^{\star +}}{\varrho^- /\varrho^{\star -}} 
= 
        j \; 
        \left[ \ln\frac{l^+}{l^-} \; 
        + \ln\frac{\varrho^{\star +}}{\varrho^{\star -}} \right] 
\equiv 
        \frac{e \, \Pi^{(+/-)}}{T} \; j , 
\]  
where $\Pi^{(+/-)}$ defines he mutual Peltier coefficient of the two materials. 
It characterizes the amount of Peltier heat produced per unit electric current, 
and is the difference of the material Peltier coefficients 
[cf.~\eq{Pi}]
\beq  
   \Pi^{(+/-)} = \Pi^+ - \Pi^- . 
 \label{eq:Pi12} 
\eeq
as also found in thermodynamics. 

\subsection{The Seebeck effect}

The Seebeck effect is observed in a thermally isolated system, where the junction 
is kept at a temperature $T_j$ different from the temperature $T_l$ prescribed 
at the leads, i.e., for the multibaker we demand 
      $T_{-M'} = T_M = T_l$ 
and   $T_{-1} = T_0 = T_j$. 
This setup corresponds to a non-uniform temperature field, and, due to this, 
also to gradients in the electro-chemical potential $\mu$. 
Because of the difference in the material properties, these gradients can add 
up to a net potential drop between the leads, even if both leads 
are kept at the same temperature and there is no particle current. 
This follows immediately from the formal definition 
\cite{GM} of the particle current in its discrete version:  
\beq 
   j_m = - \frac{\sigma_{el}}{e^2} \left[ 
             \frac{\mu_{m+1}-\mu_m}{a} \,+\, e\alpha \, 
             \frac{(T_{m+1}-T_m)}{a}   \right] ,
\label{eq:SeeJm}
\eeq 
where 
$\sigma_{el}$ is the conductivity, 
$e$ the electric charge, and 
$\alpha$ the Seebeck coefficient of the material. 
In the considered system, we then have for vanishing current 
\bega
        \mu_{-M'} - \mu_M & = &         \mu_{-M'} - \mu_{-1} + \mu_{-1} - \mu_0 + \mu_0 - \mu_M 
\nonumber \\ 
        & \approx & -  e \alpha^- \; (T_{-M'} - T_{-1}) + \mu_{-1} - \mu_0 
                        -  e \alpha^+ \; (T_{0} - T_{M})
= e (\alpha^+ - \alpha^-) \; (T_l - T_j) + \mu_{-1} - \mu_0  .
\ega
Here we have assumed the Seebeck coefficients to be 
approximately constant in the two materials. 
The macroscopic limit implies 
$ \mu_{-1}=\mu_0 $, and we obtain for the mutual Seebeck coefficient
of the two materials \cite{AM} 
\beq 
        \alpha^{(+/-)} \equiv 
        \frac{ \mu_{-M'} - \mu_M }{e (T_l - T_j) } = \alpha^+ -
        \alpha^- . 
\eeq 
It characterizes the strength of the potential drop 
$\mu_{-M'} - \mu_M$ between the leads induced by the 
temperature diffence $T_l - T_j$ between the leads and the junction. 

An expression for $\alpha^{(+/-)}$ can be determined for the
multibaker by rewriting the expression \eq{j20} for the current in the
form \eq{SeeJm}. 
Taking immediately the macroscopic limit and observing that the
electro-chemical potential can be split into a chemical part $\mu_c$
and a part $e\phi$ due to the external electric field $E\equiv-\6_x
\phi$, one obtains 
\bega 
        j 
=  - \frac{\sigma_{el}}{e^2} 
        \left[ \6_x (\mu_c + e\phi) + e \alpha \, \6_x T  \right] 
=  \frac{\sigma_{el} \, E}{e} - \frac{\sigma_{el}}{e^2} 
        \left[ \6_\vrho\mu_c \, \6_x\vrho  + \6_T\mu_c \, \6_x T  
                                           + e \alpha \, \6_x T  \right] 
=       v \vrho - D \, \6_x\vrho . 
\ega  
Here 
\begin{mathletters} 
\bega 
        v & = &  \frac{\sigma_{el} \, E}{e \, \vrho} ,
\\ 
   D      & = &  \frac{\sigma_{el}}{e^2} \; \6_\vrho\mu_c ,
\\ 
\label{eq:mu}
   e \alpha & = & - \6_T \mu_c . 
\ega 
\end{mathletters} 
By the first two equations we recover well-known relations 
from thermodynamics \cite{GM}. 
Eq.~\eq{mu} provides us with a relation for the Seebeck coefficient. 
Since the equation of state of the ``multibaker gas'' is that of a classical ideal gas 
\cite{MTV00} 
\[
\mu_c^\pm= (\gamma +1 ) T + T \ln \left( 
                           \frac{\vrho^\pm T^{-\gamma}}{\vrho^{*\pm}} \right) ,
\]
one obtains
\[ 
   e \alpha^{(+/-)} 
= 
e(\alpha^+ - \alpha^- )
= 
   e \; \ln\frac{\varrho^+ T_l^{-\gamma}}{\varrho^{\star+}} 
-  e \; \ln\frac{\varrho^- T_l^{-\gamma}}{\varrho^{\star-}} 
=  
   e \;  \left[ \ln\frac{l^+}{l^-} \; 
   + \ln\frac{\varrho^{\star+}}{\varrho^*} \right] 
= 
   \frac{e \, \Pi^{(+/-)}}{T_l} , 
\] 
where  (\ref{eq:Pi12})  was used in the last step. 
This comparison expresses the validity of the Onsager relation 
   $\Pi^{(+/-)}= \alpha^{(+/-)} T$ 
for this class of models.

\section{Discussion} 

In this paper we have described the local and global transport
properties of multibakers with a density and an energy dynamics. 
This class of maps makes an analytical modelling of 
transport processes by a deterministic chaotic dynamics possible, and
admits a macroscopic description consistent with various aspects of 
irreversible thermodynamics. 
The macroscopic description comprises the time evolution of the average 
density and the kinetic energy in small regions of the physical space (the cells
of the multibaker).
The former density is interpreted as the particle density, and the latter as a 
temperature field. 
The averages in the small regions are in the spirit of local
thermodynamic equilibrium, and the continuum description of
thermodynamics arises in a
macroscopic limit where the spatial resolution of the tranport process
is small compared to the system size (or any other relevant
macroscopic length), and where a discrete time-scale used in the
definition of the dynamics is much smaller than macroscopic time
scales. 

The relevant concept of entropy for multibakers is the Gibbs
entropy defined with respect to the average density in the cells 
normalized by a temperature-dependent reference density. 
It is called the coarse-grained entropy. 
Based on an information-theoretic interpretation of the entropy, a
local entropy balance can be derived, which in the 
macroscopic limit can be fully consistent with irreversible thermodynamics. 
This agreement holds provided that 
(i) a particular choice of local phase-space contraction and expansion 
rates is incorporated in the time evolution of the density, which 
we identified as a time-reversible evolution of 
the mapping in previous work \cite{VTB97,VTB98,BTV98}, 
(ii) the density in the entropy is normalized by a reference density 
with a power-law dependence on the average kinetic energy in the cell, and 
(iii) appropriate source terms are incorporated in the evolution 
equations of the kinetic-energy field. 
No meaningful macroscopic description can be found for multibakers
with other choices of the phase-space contraction factors. 
Modification of the source terms leads to additional contributions in
the local entropy balance, which are interpreted as local entropy
fluxes into a thermostat. 
In particular, for vanishing source terms one can mimic a transport
process in a system with a spatially uniform temperature, i.e., one
obtains a setting reminiscent of NEMD simulations of transport
processes. 

Once the connection between the deterministic dynamics of the
multibaker and the corresponding local thermodynamic relations is
established, one can apply it to discuss transport in different
macroscopic settings. 
A number of models with periodic boundary conditions were discussed in 
order to shed light on the global entropy balance in such systems. 
We find that, up to a trivial factor, the average phase-space contraction 
amounts to the entropy flux to the environment. 
This supports an earlier heuristic argument of Ruelle \cite{Ru} and others 
\cite{CELS,ECM,GC}, who connected the phase-space contraction to the irreversible 
entropy production in a steady state. 
In contrast to the claims of some of the latter authors (cf.~for instance 
\cite{RC99}) the connection between the irreversible entropy production and 
the phase-space contraction rate breaks down away from stationarity. 
In fact, the contraction rate is still connected to the entropy flux
in that situation, but the flux is no longer related to the rate of 
irreversible entropy production. 
This was shown 
(a) for multibakers with a uniform thermostatting (Model II),
i.e., for models reminiscent of NEMD algorithms, 
(b) for systems where the driving and thermostatting is applied in a
macroscopically small region of the system (Model III), thus giving
rise to sustained density gradients, and 
(c) systems with a uniform external field and localized thermostatting
(Model IV).  
The former two models have constant temperature fields, while the
latter one supports a temperature profile with a
discontinuity in the first derivative at the positon of thermostatting. 
As expected from the existence of the local entropy balance, the
results are fully consistent with the corresponding
thermodynamic description of the transport process. 
They suggest an interesting conclusion on modelling transport 
in bulk systems by isothermal NEMD simulations: 
These methods are valid in an approximation where the considered
volume is sufficiently small to neglect density 
and temperature gradients. 
In steady states, they are equivalent to models, where the currents 
are the same, but thermostatting is only applied at the boundaries of
a macroscopic system. 
Since even state of the art simulations can hardly cope with more than 
$10^9$ particles, i.e., with integration volumes larger than about 
$\mu$m$^3$, this approximation seems to be well-justified in numerical 
studies. 
On the other hand, this assumptions should be kept in mind when isothermal 
NEMD modelling is taken as basis of theoretical studies of
transport processes (cf.~for instance \cite{CELS,ECM,GC,Gal}). 

To further demonstrate the use of multibakers with density and 
energy fields, we also discussed thermoelectric cross effects. 
The description of the transport properties requires in that case
information on the equation of state, since the Seebeck effect is
defined in terms of differences of chemical potentials. 
In previous work \cite{MTV99,MTV00} it was shown that the classical 
ideal-gas equation holds for multibakers. 
This is meaningful since the time evolution of the multibaker can be
considered as the one of particles with phase-space coordinates
$(x,p)$, which only interact by a (weak) mean-field like coupling
manifested in a dependence of the local parameters on the average 
densities. 
With this input the Peltier and Seebeck effect were modelled and the 
Onsager relation, connecting their respective transport coefficients, 
was derived. 
The validity of this relation for multibakers is not a trivial result. 
It heavily relies on the choices (i)--(iii) to find an entropy
balance consistent with irreversible thermodynamics. 

Summarizing, we demonstrated that multibakers establish a straightforward 
modelling of various transport phenomena by deterministic, chaotic
dynamics. 
They give insight in the general structure of such models by explicit
analytical calculations. 
This was demonstrated by discussions of thermoelectric cross effects,
and of the relation between the average phase-space contraction,
entropy fluxes and the rate of irreversible entropy production. 

\acknowledgements 

This work is dedicated to Prof.~Gregoire Nicolis on occasion of his 60th birthday. 
Our discussions at the {\em Centre of Complex Systems and Nonlinear Phenomena\/} 
helped us a lot, when we were starting this project in the summer of 1998. 
We are also grateful to J.R.~Dorfman, B.~Fogarassy, J.~Hajdu, G.~Tichy and H.~Posch 
for enlightening discussions. 
The paper was completed during a joint stay at the {\em Max-Planck Institute for 
the Physics of Complex Systems\/}, which we gratefully acknowledge together with 
support from the Hungarian Science Foundation (OTKA T17493, T19483) and the 
TMR-network {\em Spatially extended dynamics\/}. 


\end{document}